\documentclass[journal,twoside,web]{ieeecolor}
\usepackage{generic}
\usepackage{stfloats}
\usepackage{multirow}
\usepackage{amsmath,amssymb,amsfonts}
\usepackage{algorithmic}
\usepackage{graphicx}
\usepackage{algorithm,algorithmic}
\usepackage{hyperref}
\usepackage{booktabs}
\usepackage{xcolor}  
\usepackage{ulem}    
\usepackage{cite}           
\usepackage[section]{placeins} 
\usepackage[section]{placeins} 
\usepackage{booktabs}

\usepackage{textcomp} 
\def\BibTeX{{\rm B\kern-.05em{\sc i\kern-.025em b}\kern-.08em
    T\kern-.1667em\lower.7ex\hbox{E}\kern-.125emX}}
\begin{document}
\title{SPIDER: Structure-Preferential Implicit Deep Network for Biplanar X-ray Reconstruction}
\author{Tianqi~Yu, Xuanyu~Tian, Jiawen~Yang, Dongming~He, Jingyi~YU, Xudong~Wang ,and Yuyao~Zhang
\thanks{}
}
\maketitle

\begin{abstract}
Biplanar X-ray imaging is widely used in health screening, postoperative rehabilitation evaluation of orthopedic diseases, and injury surgery due to its rapid acquisition, low radiation dose, and straightforward setup. However, 3D volume reconstruction from only two orthogonal projections represents a profoundly ill-posed inverse problem, owing to the intrinsic lack of depth information and irreducible ambiguities in soft-tissue visualization. Some existing methods can reconstruct skeletal structures and Computed Tomography (CT) volumes, they often yield incomplete bone geometry, imprecise tissue boundaries, and a lack of anatomical realism, thereby limiting their clinical utility in scenarios such as surgical planning and postoperative assessment.

In this study, we introduce SPIDER, a novel supervised framework designed to reconstruct CT volumes from biplanar X-ray images. SPIDER incorporates tissue structure as prior (e.g., anatomical segmentation) into an implicit neural representation decoder in the form of joint supervision through a unified encoder-decoder architecture. This design enables the model to jointly learn image intensities and anatomical structures in a pixel-aligned fashion.

To address the challenges posed by sparse input and structural ambiguity, SPIDER directly embeds anatomical constraints into the reconstruction process, thereby enhancing structural continuity and reducing soft-tissue artifacts. We conduct comprehensive experiments on clinical head CT datasets and show that SPIDER generates anatomically accurate reconstructions from only two projections. Furthermore, our approach demonstrates strong potential in downstream segmentation tasks, underscoring its utility in personalized treatment planning and image-guided surgical navigation.

\end{abstract}


\section{Introduction}
\label{sec:introduction}
\IEEEPARstart{X}{-ray} imaging remains one of the most widely utilized modalities in clinical diagnostics due to its rapid acquisition speed and significantly lower radiation dose compared to other imaging techniques. However, its inherent limitation—lack of depth information—renders X-ray-based diagnosis heavily reliant on physician expertise and interpretative skill \cite{kalender2011computed}. In contrast, Computed Tomography (CT) reconstructs volumetric representations from densely sampled X-ray projections, delivering highly detailed and anatomically accurate 3D visualizations \cite{hsieh2003computed}. Despite its advantages, CT involves several limitations, including elevated radiation exposure, reduced sampling efficiency, and logistical constraints that hinder its deployment in intraoperative settings or repeated scanning scenarios \cite{brenner2007computed}.

Biplanar X-ray imaging offers a compelling intermediate solution by capturing orthogonal projections, thereby introducing partial spatial context while preserving the advantages of low-dose and rapid acquisition. The additional viewpoint helps mitigate depth ambiguity inherent in single-view X-rays, enabling the reconstruction of 3D structures from two 2D images. Recent research has explored the feasibility of using biplanar X-rays for tasks such as anatomical registration or volume reconstruction via parametric modeling \cite{loisel2023three,yang20242d,kadoury2008statistical,humbert20093d}. However, despite the supplementary perspective, the persistent lack of explicit depth cues and frequent soft-tissue overlap render the reconstruction task a fundamentally ill-posed inverse problem.

The advent of deep learning, coupled with the increasing availability of large-scale medical imaging datasets \cite{wasserthal2023totalsegmentator,liu2021deep, sekuboyina2021verse,lin2023rsna,armato2011lung}, has fueled the development of data-driven approaches to tackle this reconstruction challenge. Supervised models—particularly those leveraging convolutional neural networks (CNNs) have been widely adopted to learn direct mappings from X-ray projections to 3D volumes \cite{X-CTRSNet,X-ray2CTNet,X2ctgan,Xtransct,CCX}. 
While these methods demonstrate the potential to reconstruct volumes from biplanar X-rays, the quality of the resulting reconstructions still leaves room for improvement. Some deeper 3D-CNN networks also have extremely high storage requirements, which makes the reconstruction of high-resolution volumes highly dependent on the equipment.

Some GAN (generative adversarial networks) based works have demonstrated their ability to obtain 3D volumes from biplanar X-rays due to their powerful model representation capabilities. Ying et al. proposed X2CTGAN~\cite{X2ctgan}, introduced adversarial loss and perspective consistency mechanisms to reconstruct 3D chest CT volumes from biplanar images. However, research indicates that accuracy and realism are a trade-off in GANs~\cite{wang2018esrgan}, as GANs excel at optimising perceptual consistency rather than precise numerical or structural reconstruction. This results in the 3D volumes reconstructed by X2CTGAN appearing realistic but exhibiting limited anatomical accuracy.

To address these limitations, recent studies have proposed the integration oftion oftion of of supervised learning with implicit neural representation (INR) frameworks \cite{DIFNet, kyung2023perspective, X2ctCNN}. These approaches leverage the feature extraction strengths of CNNs in the image domain alongside the expressive power and continuity of INRs, employing a 2D image encoder paired with an implicit decoder to perform point-wise reconstruction of CT volumes. This design not only efficiently extracts features from both planes while reducing computational cost, but also bypasses the need for global texture matching, resulting in higher fidelity compared to GAN-based methods. However, since INRs aim to fit smooth functions over the entire volume, the inherent smoothness bias of neural networks encourages continuity, making it challenging to reconstruct sharp boundaries—such as interfaces between bones and soft tissues, joints, or organ edges—without additional guidance or constraints. In the context of biplanar X-ray applications, accurate structural information is more critical than texture details \cite{borzabadi2012orthodontic, schmalbach2010anterior, zuniga2016updates}, especially for tasks like postoperative assessment and orthognathic surgery planning. Therefore, it is essential to guide the network to prioritize reconstructing structures with well-defined boundaries, even when operating with extremely sparse biplanar input data.

In this paper, we propose SPIDER: Structure-Preferential Implicit DEep network for biplanar x-ray Reconstruction, proposes a new method of structural augmentation constraints applied to implicit representation models. Unlike previous approaches that struggle to balance skeletal and soft tissue representation, SPIDER's unique strength lies in its ability to jointly optimize the reconstruction of image intensity and structural features through a shared decoder. This innovative design enables precise learning of pixel-level correspondences between appearance and geometry. By embedding structural priors, such as segmentations, to guide network reconstruction 3D volumes with clearly defined boundariesour. SPIDER stands out by achieving highly anatomically consistent reconstructions. SPIDER stands out by achieving highly anatomically consistent reconstructions. It sharpens the delineation of critical  structures with unprecedented clarity, outperforming existing techniques in meeting the dual demands of clinical imaging.


The main contributions of this work are as follows: \begin{enumerate} \item[(1)] We propose \textbf{SPIDER}, the first framework, to the best of our knowledge, that explicitly incorporates structural priors into biplanar X-ray reconstruction. Our method ensures anatomical consistency by enhancing skeletal continuity while preserving soft tissue details. \item[(2)] We introduce a novel structure-preferential supervised implicit representation paradigm, in which structural and image features are predicted jointly via a shared decoder, enabling improved alignment between appearance and geometry. \item[(3)] We demonstrate the effectiveness of our approach on real clinical data, achieving structurally complete and visually accurate reconstructions. Our method also shows potential for downstream tasks such as segmentation on simulated datasets. \end{enumerate}

\section{PRELIMINARIES}
\label{sec:preliminaries}
\subsection{Acquisition and Reconstruction of Biplanar X-ray}

In X-ray imaging, the intensity \( I(u,v) \) measured at a detector pixel \((u,v)\) is governed by the Beer--Lambert law:
\begin{equation}
I(u,v) = I_0(u,v) \exp\left( - \int_{L(u,v)} \mu(x,y,z) \, \mathrm{d}l \right),
\end{equation}
where \( I_0(u,v) \) is the known incident intensity, \( \mu(x,y,z) \) is the spatially-varying linear attenuation coefficient, and the integral is taken along the X-ray path \( L(u,v) \). Taking the negative logarithm yields a linearized measurement:
\begin{equation}
p(u,v) = -\log \frac{I(u,v)}{I_0(u,v)} = \int_{L(u,v)} \mu(x,y,z) \, \mathrm{d}l.
\end{equation}

When a full set of projections is acquired (i.e., under dense angular sampling), the CT reconstruction can be formulated as a linear inverse problem. Let the continuous attenuation field be discretized into a vector \( \mathbf{x} \in \mathbb{R}^N \), and the collection of all logarithmic projection measurements be denoted by \( \mathbf{y} \in \mathbb{R}^M \). Then the forward model is expressed as:
\begin{equation}
\mathbf{y} = \mathbf{A} \mathbf{x} + \mathbf{b},
\end{equation}
where \( \mathbf{A} \in \mathbb{R}^{M \times N} \) is the system matrix whose rows represent the discrete line integrals along the X-ray paths, and \( \mathbf{b} \) accounts for measurement noise and modeling errors. Even in the dense case, the reconstruction of \( \mathbf{x} \) from \( \mathbf{y} \) is an ill-posed inverse problem, requiring appropriate regularization techniques.

In practical applications where only two orthogonal X-ray projections are available (commonly referred to as biplanar X-ray imaging), the reconstruction problem becomes significantly more challenging. Denote the measurements from the two views by \( \mathbf{y}_1  \) and \( \mathbf{y}_2 \in \mathbb{R}^{M} \), with corresponding system matrices \( \mathbf{A}_1  \) and \( \mathbf{A}_2 \in \mathbb{R}^{M \times N} \). The combined measurement model is then given by:
\begin{equation}
\begin{pmatrix}
\mathbf{y}_1 \\
\mathbf{y}_2
\end{pmatrix}
=
\begin{pmatrix}
\mathbf{A}_1 \\
\mathbf{A}_2
\end{pmatrix}
\mathbf{x}
+
\begin{pmatrix}
\mathbf{b}_1 \\
\mathbf{b}_2
\end{pmatrix}.
\end{equation}
Since the total number of measurements \( M = M_1 + M_2 \) is typically much smaller than the number of unknowns \( N \) (i.e. \( M \ll N \)), the combined system matrix
\begin{equation}
\mathbf{A}_{\text{bp}} = \begin{pmatrix} \mathbf{A}_1 \\ \mathbf{A}_2 \end{pmatrix} \in \mathbb{R}^{M \times N}
\end{equation}
is severely rank-deficient. In other words,
\begin{equation}
\mathrm{rank}(\mathbf{A}_{\text{bp}}) \ll N,
\end{equation}
which implies the existence of an infinite number of solutions \( \mathbf{x}' \neq \mathbf{x} \) satisfying
\begin{equation}
\mathbf{A}_{\text{bp}} \mathbf{x}' = \mathbf{A}_{\text{bp}} \mathbf{x}.
\label{eq:undetermined}
\end{equation}
Thus, the biplanar reconstruction problem is an underdetermined and ill-posed inverse problem, necessitating the introduction of additional constraints or regularization to obtain a unique and stable solution.

\begin{figure*}[t]
    \centering
    \includegraphics[width=1\linewidth]{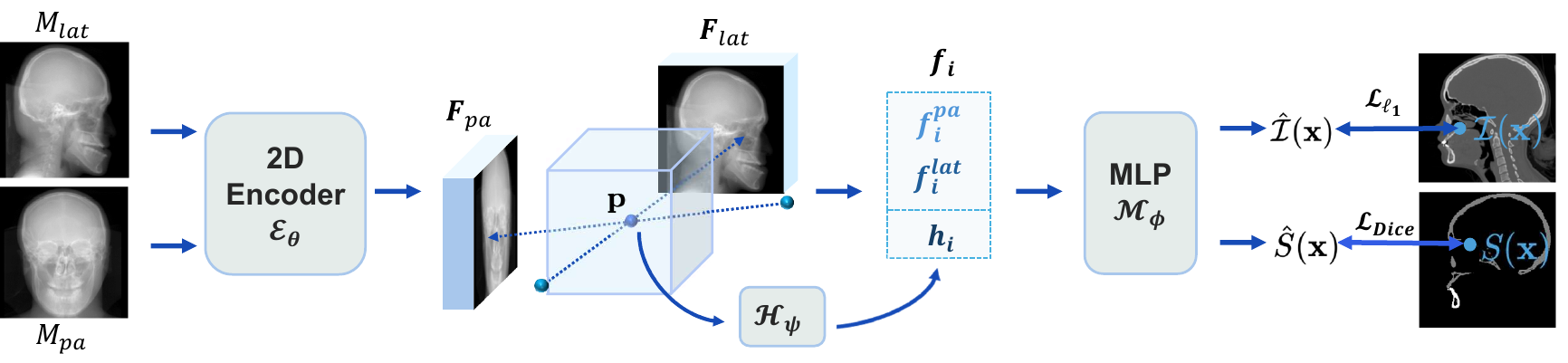}
    \caption{Overview of the proposed SPIDER.
(a) Given biplanar X-rays $M_{pa},M_{lat} $, we employ a CNN-based encoder $\mathcal{E}_{\theta}$ to generate view-specific feature maps. Meanwhile, we get the encoded spatial coordinates $h_i$ via hash encoding $\mathcal{H}_\psi$.
(b) For an arbitrary spatial point $\mathbf{p}$, we utilize the ray trajectory to project $\mathbf{p}$ onto both feature maps to obtain the view-aligned features ${f_i^{pa}, f_i^{lat} }$. 
These features are concatenated with $h_i$ to form a feature vector containing 3D spatial information.  
(c) Finally, the combined feature is decoded by an MLP network $\mathcal{M}_{\phi}$ to simultaneously predict the intensity and segmentation label at point $\mathbf{p}$.}
\label{fig:enter-label}
\end{figure*}

\subsection{Image-Conditioned Implicit Neural Representation}

Implicit Neural Representations (INRs) have emerged as a powerful alternative to traditional grid-based representations by modeling signals as continuous functions parameterized by neural networks. Formally, a signal \( f : \mathbb{R}^n \rightarrow \mathbb{R}^m \) is approximated by a neural function \( \Phi_\theta \), such that:
\begin{equation}
f(\mathbf{x}) \approx \Phi_\theta(\mathbf{x}),
\end{equation}
where \( \mathbf{x} \in \mathbb{R}^n \) denotes a spatial coordinate, and \( \Phi_\theta(\mathbf{x}) \) represents the corresponding signal value (e.g., CT intensity or segmentation label). To enhance the network's ability to capture fine-grained details, input coordinates are typically transformed using high-frequency positional encodings \( \gamma(\mathbf{x}) \), such as Fourier features, or trainable encodings like hash embeddings.

Recent extensions to the standard INR framework have incorporated external conditioning information, typically derived from image encoders. In this formulation, a feature vector \( \mathbf{f}(\mathbf{x}) \) is extracted from an image encoder at the spatial location corresponding to \( \mathbf{x} \), and then fused with the positional encoding \( \gamma(\mathbf{x}) \). This fusion can be achieved via concatenation, coordinate-based querying, or cross-attention mechanisms~\cite{lee2023locality,ekanayake2025seco,zhang2024attention}, resulting in a conditioned input to the implicit decoder:
\begin{equation}
\mathcal{V}(\mathbf{x}) = \Phi_\theta([\mathbf{f}(\mathbf{x}), \gamma(\mathbf{x})]),
\end{equation}
where \( \mathcal{V}(\mathbf{x}) \) denotes the predicted signal value. This image-conditioned formulation allows the network to leverage both global context from input images and the local spatial structure inherent in the coordinate representation, leading to more accurate and perceptually coherent reconstructions.

\section{METHODS}

\label{sec:methods}

In this work, we first encode a biplanar Xray using a 2D encoder (Section\ref{subsec:enc}) and query the feature vectors at the corresponding positions of the predicted points in space in the context of the actual acquisition environment, combine them with multiresolution hash coding into an implicit neural representation-based decoder (Section\ref{subsec:sample}), and output both the predicted intensity information and the carriers containing the structural information through a shared decoder (Section\ref{subsec:predict}). With this special structural constraint, the implicit neural representation-based reconstruction algorithm makes the reconstructed volume have better structural information.
\subsection{X-ray Feature Encoding}
\label{subsec:enc}
We employ a UNet~\cite{ronneberger2015u} encoder $\mathcal{E}_\theta$ to extract view-specific features from X-rays $M_{pa}$ and $M_{lat} \in \mathbb{R}^{H \times W}$, where $M_{pa}$ and $M_{lat}$ represent the Posterior-Anterior and Lateral X-rays, respectively, and $H$ and $W$ denote their spatial dimensions. The UNet architecture is chosen due to its powerful multi-scale feature extraction capability, which is achieved through a series of convolutional layers combined with down-sampling operations that capture both local texture and global context. Additionally, the skip connections inherent to UNet facilitate the preservation of fine-grained spatial information that is critical for subsequent tasks such as segmentation and registration.

Specifically, the two orthogonal X-ray views are processed through a shared 2D UNet encoder, ensuring that both images are mapped into a consistent feature space. This process can be represented as follows:
\begin{equation}
    \mathbf{F}_\text{pa} = \mathcal{E}_\theta(M_\text{pa}), \quad 
    \mathbf{F}_\text{lat} = \mathcal{E}_\theta(M_\text{lat}),
\end{equation}
where $\left \{\mathbf{F}_\text{pa}, \mathbf{F}_\text{lat} \right \} \subset \mathbb{R}^{H \times W \times C}$ denotes the extracted multi-channel feature maps and $C$ is the number of feature channels. By sharing the encoder parameters, the model enforces a consistent representation across different views.

\subsection{Sampling Module}
\label{subsec:sample}
To effectively simulate the acquisition process of dual-plane X-rays and integrate both spatial and view-specific information, we propose a projection-guided feature sampling strategy. Given a set of 3D query points $\{\mathbf{x}_i \in \mathbb{R}^3\}_{i=1}^N$ sampled in the canonical CT space, we back-project each point onto the two imaging planes corresponding to the Posterior-Anterior (PA) and Lateral (LAT) X-ray views. These planes are restored to their original acquisition poses, denoted by extrinsic parameters $\mathcal{T}_\text{pa}$ and $\mathcal{T}_\text{lat}$, respectively.

The projection of a 3D point $\mathbf{x}_i$ onto each X-ray plane is computed via a perspective projection:
\begin{equation}
    \mathbf{p}_i^\text{pa} = \Pi(\mathcal{T}_\text{pa}, \mathbf{x}_i), \quad 
    \mathbf{p}_i^\text{lat} = \Pi(\mathcal{T}_\text{lat}, \mathbf{x}_i),
\end{equation}
where $\Pi(\cdot)$ denotes the X-Ray projection function that maps 3D coordinates to 2D image space. Using bilinear interpolation at these projected positions, we retrieve corresponding feature vectors from the extracted feature maps $\mathbf{F}_\text{pa}$ and $\mathbf{F}_\text{lat}$:
\begin{equation}
    \mathbf{f}_i^\text{pa} = \text{Interp}(\mathbf{F}_\text{pa}, \mathbf{p}_i^\text{pa}), \quad
    \mathbf{f}_i^\text{lat} = \text{Interp}(\mathbf{F}_\text{lat}, \mathbf{p}_i^\text{lat}).
\end{equation}

To efficiently encode spatial coordinates, we adopt the multi-resolution hash encoding strategy proposed in Instant-NGP~\cite{muller2022instant}. This approach maps continuous coordinates into a series of resolution levels using sparse learnable hash tables, enabling high-fidelity representation with a compact memory footprint.

Let $\mathbf{x} \in [0, 1]^3$ be a point in normalized 3D space. The multi-resolution hash encoder constructs $L$ resolution levels. At each level $l$, the space is divided into a $T_l \times T_l \times T_l$ grid. Each grid cell corner is associated with a learnable feature vector, retrieved via a hash function.

At level $l$, the input coordinate is scaled to the grid: \begin{equation} \mathbf{u}^{(l)} = \mathbf{x} \cdot T_l \end{equation} We determine the 8 grid vertices surrounding $\mathbf{u}^{(l)}$ and interpolate their features using trilinear interpolation. Each vertex at integer coordinate $\mathbf{i} = (i_x, i_y, i_z)$ is mapped to a feature vector via a hash function: \begin{equation} \text{hash}(\mathbf{i}) = \left( (i_x \cdot p_1) \oplus (i_y \cdot p_2) \oplus (i_z \cdot p_3) \right) \bmod T_\text{max} \end{equation} where $p_1$, $p_2$, and $p_3$ are large, fixed prime numbers, and $\oplus$ denotes the bitwise XOR operator. $T_\text{max}$ is the maximum number of entries in the hash table for each level.

Let $\mathcal{H}^{(l)}: \mathbb{Z}^3 \rightarrow \mathbb{R}^F$ be the hash table at level $l$, with $F$ being the feature dimension per grid point. The interpolated feature vector at level $l$ is then computed as: \begin{equation} \gamma^{(l)}(\mathbf{x}) = \sum_{j=1}^{8} w_j \cdot \mathcal{H}^{(l)}\left( \text{hash}(\mathbf{i}_j) \right) \end{equation} where $w_j$ are the trilinear interpolation weights for the 8 neighboring vertices $\mathbf{i}_j$.

The final multi-resolution encoding is formed by concatenating the encodings across all levels: \begin{equation} \gamma(\mathbf{x}) = \text{concat}\left[ \gamma^{(1)}(\mathbf{x}), \gamma^{(2)}(\mathbf{x}), \dots, \gamma^{(L)}(\mathbf{x}) \right] \in \mathbb{R}^{L \cdot F} \end{equation}

To encode the geometric information of each sampled point, we additionally apply a multi-resolution hash encoding $\mathcal{H}_\phi$ to its 3D coordinate:
\begin{equation}
    \mathbf{h}_i = \mathcal{H}_\psi(\mathbf{x}_i),
\end{equation}
where $\mathcal{H}_\psi: \mathbb{R}^3 \rightarrow \mathbb{R}^{C_h}$ maps the point to a compact, high-dimensional representation. Finally, the full feature vector for each point $\mathbf{x}_i$ is formed by concatenating view-specific and spatial encodings:
\begin{equation}
    \mathbf{f}_i = \text{Concat}(\mathbf{f}_i^\text{pa}, \mathbf{f}_i^\text{lat}, \mathbf{h}_i) \in \mathbb{R}^{2C + C_h}.
\end{equation}

This feature embedding captures both appearance cues from dual-view X-rays and precise spatial context, providing rich supervision for downstream volumetric reconstruction or segmentation tasks.

\subsection{Joint Intensity and Structural Predictions}
\label{subsec:predict}
To jointly model volumetric intensity and structural information, we employ a unified implicit neural representation. The network takes a feature vector, sampled from feature maps generated by the encoder and position coordinate codes, as input and simultaneously predicts the corresponding intensity and semantic structure.

Given a spatial coordinate $\mathbf{x} = (x, y, z)$, We can obtain the features $f_i$ corresponding to this point by the previous sampling module. The feature is then passed through a decoder network $\mathcal{M}_\phi$:
\begin{equation}
    \mathbf{o}(\mathbf{x}) = \mathcal{M}_\phi\left( f_i \right) \in \mathbb{R}^{C + 1}
\end{equation}
where $\mathbf{o}(\mathbf{x})$ denotes the output of the network, and $C$ is the number of structural classes including background.

The output vector $\mathbf{o}(\mathbf{x})$ is divided into:
\begin{itemize}
    \item \textbf{Intensity prediction:} the first channel represents the reconstructed intensity:
    \begin{equation}
        \hat{\mathcal{I}}(\mathbf{x}) = \mathbf{o}_0(\mathbf{x})
    \end{equation}
    
    \item \textbf{Structural prediction:} the remaining $C$ channels correspond to the logits for semantic class prediction:
    \begin{equation}
        \hat{\mathbf{s}}(\mathbf{x}) = \left[ \mathbf{o}_1(\mathbf{x}), \dots, \mathbf{o}_C(\mathbf{x}) \right] \in \mathbb{R}^C
    \end{equation}
    
    For segmentation prediction,  a softmax function is applied to obtain class probabilities :
     \begin{equation}
        \mathbf{p}(\mathbf{x}) = \text{softmax}\left( \hat{\mathbf{s}}_{m:n}(\mathbf{x}) \right)
    \end{equation}
    The final predicted class is obtained as:
    \begin{equation}
        \hat{\mathcal{S}}(\mathbf{x}) = \arg\max_{c \in \{m, \dots, n\}} \, \mathbf{p}_c(\mathbf{x})
    \end{equation}
\end{itemize}

This architecture enables the model to reconstruct both intensity and semantic structures from a shared continuous volumetric representation.

\subsection{Network Optimization}

To jointly optimize the implicit neural representation for intensity reconstruction, structural segmentation learning, we employ a composite loss function that integrates multiple objectives. Specifically, we define three loss components: intensity reconstruction loss and segmentation loss.

\paragraph{Intensity Reconstruction Loss}  
For the predicted intensity $\hat{I}(\mathbf{x})$ and the ground truth intensity $I(\mathbf{x})$, we use the $L_1$ loss to preserve fine-grained details and mitigate over-smoothing:
\begin{equation}
    \mathcal{L}_{\text{int}} = \mathbb{E}_{\mathbf{x} \in \Omega} \left[ \left| \hat{\mathcal{I}}(\mathbf{x}) - \mathcal{I}(\mathbf{x}) \right| \right]
\end{equation}
where $\Omega$ denotes the domain of sampled spatial points.


\paragraph{Segmentation Loss}  
For the semantic segmentation task, we use the Dice loss~\cite{milletari2016v} to handle class imbalance and improve structural consistency. Given the predicted class probabilities $\mathbf{p}(\mathbf{x}) \in \mathbb{R}^C$ and the one-hot encoded ground truth label $\mathbf{y}(\mathbf{x}) \in \{0,1\}^C$, the Dice loss is computed as:
\begin{equation}
    \mathcal{L}_{\text{seg}} = 1 - \frac{2 \sum_{c=1}^{C} \sum_{\mathbf{x} \in \Omega} \mathbf{p}_c(\mathbf{x}) \mathbf{y}_c(\mathbf{x}) + \epsilon}
    {\sum_{c=1}^{C} \sum_{\mathbf{x} \in \Omega} \mathbf{p}_c(\mathbf{x}) + \mathbf{y}_c(\mathbf{x}) + \epsilon}
\end{equation}
where $\epsilon$ is a small constant to ensure numerical stability.

\paragraph {Final Objective}  
The overall training objective is a weighted sum of the three loss components:
\begin{equation}
    \mathcal{L}_{\text{total}} = \lambda_{\text{int}} \mathcal{L}_{\text{int}} +   \lambda_{\text{seg}} \mathcal{L}_{\text{seg}}
\end{equation}
where $\lambda_{\text{int}}$ and $\lambda_{\text{seg}}$ are hyperparameters that control the contribution of each task to the total loss.

This joint loss formulation facilitates the concurrent learning of both appearance and geometric structure, enhancing the representation capacity of the network across multiple tasks.

\section{EXPERIMENTS}
\label{sec:experiments}

\begin{itemize} 
    \item \textbf{LUCY dataset}: The LUCY~\cite{Lucy} dataset was collected from archived medical records at the Department of Oral \& Craniomaxillofacial Surgery, Ninth People's Hospital, Shanghai Jiao Tong University School of Medicine. It includes 138 craniofacial CT scans with expert annotations for the maxilla and mandible. We randomly split the dataset into 100 training, 18 validation, and 20 test subjects.

    \item \textbf{Total Segmentator dataset}: The Total Segmentator dataset~\cite{totalsegmentator} is an open-source collection of 3D CT scans with comprehensive anatomical segmentations, covering common organs, bones, and muscles relevant to clinical practice. We selected 206 cases with similar anatomical coverage and preprocessing, splitting them into 150 training, 28 validation, and 28 test cases.
\end{itemize}

To simulate biplanar X-rays from the CT volumes, we used the Core Imaging Library (CIL)~\cite{CILp1,CILp2} under a parallel-beam projection geometry. Each scan was projected from two orthogonal views, and the resulting synthetic X-ray images were generated at a resolution of $128 \times 128$.

\subsection{Compared Methods}

We compare our proposed method against three baselines, including: \begin{itemize} \item \textbf{FDK}\cite{FDK}: A classical analytical reconstruction algorithm. \item \textbf{X2CT-GAN}\cite{X2ctgan}: A generative adversarial network trained for X-ray-to-CT volume translation. \item \textbf{PerX2CT}~\cite{kyung2023perspective}: A state-of-the-art deep learning framework that integrates perspective-aware features into an implicit neural representation. \end{itemize}

We evaluate reconstruction performance using two standard image quality metrics: Peak Signal-to-Noise Ratio (PSNR) and Structural Similarity Index Measure (SSIM)~\cite{SSIM}, comparing reconstructed volumes against the ground truth CT scans.

To further assess the clinical utility of the reconstructions, we evaluate the anatomical fidelity of reconstructed skeletal, muscular, and organ structures using a downstream segmentation task. Specifically, we apply a pretrained nnUNet~\cite{isensee2021nnu} model based on the TotalSegmentator annotations~\cite{wasserthal2023totalsegmentator}, and report the segmentation performance on representative anatomical regions (e.g., kidneys, hips, sacrum, vertebrae, psoas major, and gluteal muscles). This indirect evaluation reflects both the visibility and spatial accuracy of clinically relevant structures.

We use the following three metrics for quantitative evaluation: \begin{itemize} \item \textbf{Dice coefficient}~\cite{Dice}: Measures volumetric overlap. \item \textbf{95th percentile Hausdorff Distance (HD95)}: Evaluates boundary accuracy and robustness to outliers. \item \textbf{Chamfer Distance (CD)}: Captures overall geometric fidelity between segmented surfaces. \end{itemize}

\subsection{Implementation Details}

We train SPIDER using stochastic gradient descent (SGD), with a structural prior weighting coefficient $\lambda = 0.3$ in the total loss. The learning rate is initialized to 0.001 and decayed by a factor of 0.5 every 100 epochs. Training is conducted for 500 epochs.For the hash encoding module, we set the encoding levels $L = 11$ and feature dimensions $F = 8$. All experiments are performed on a single NVIDIA RTX 4090 GPU.

\section{Results}

\subsection{Comparison of Reconstructed CT Volume}

\begin{figure*}[ht]
    \centering
    \includegraphics[width=1\linewidth]{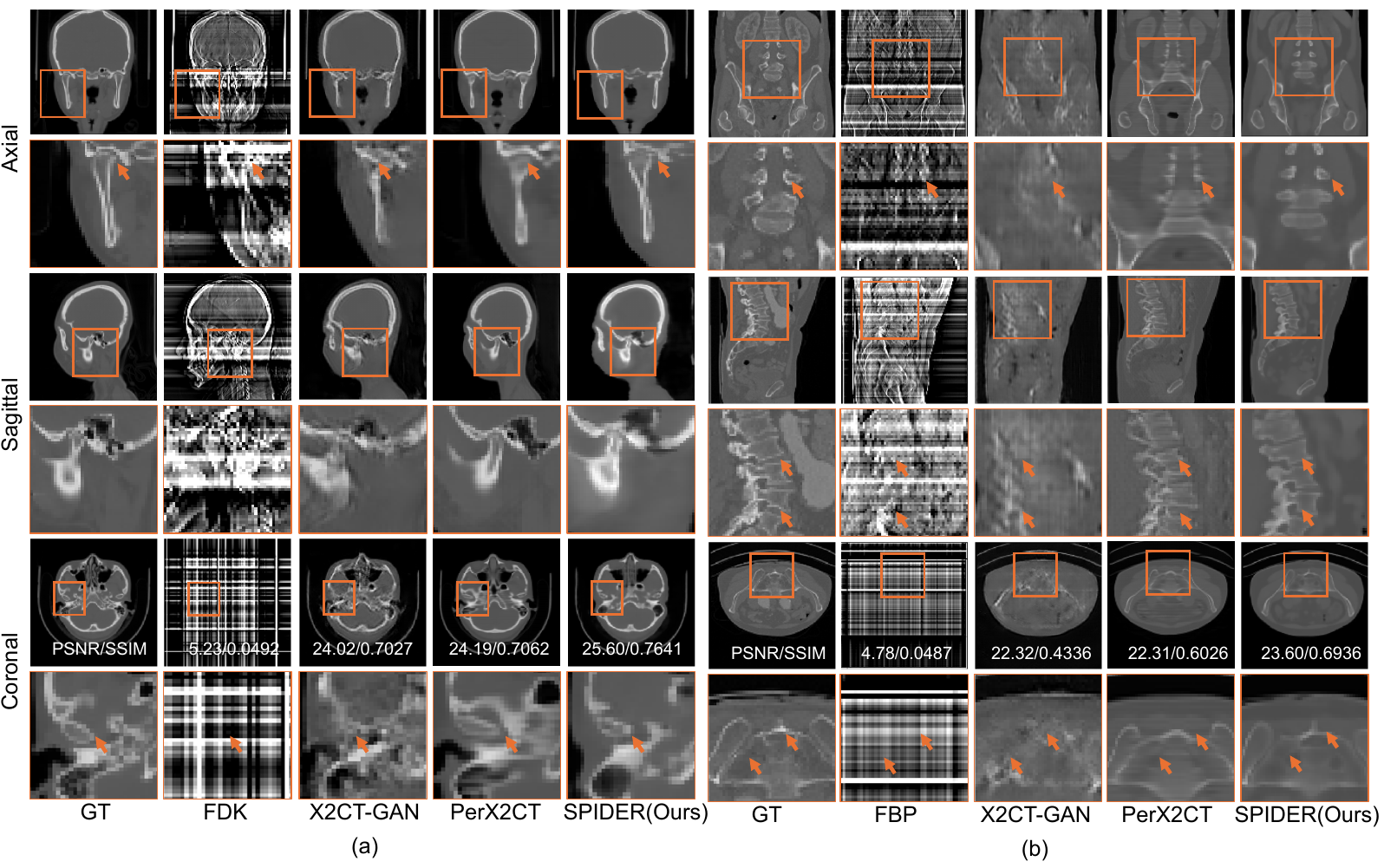}
    \caption{Comparison of CT volume reconstruction of baseline methods from biplanar
X-ray at three different views. (a) shows the result of methods in LUCY dataset.(b) shows the experimental results on the TotalSegmentator dataset. }
    \label{fig:recon}
\end{figure*}

In our experiments based on the LUCY dataset, we gave information on two types of structures, the maxilla and the mandible.
Figure~\ref{fig:recon} (a) illustrates representative 3D reconstructions from different methods in LUCY dataset. 
Biplanar X-rays do not provide enough information to reconstruct a usable volume with conventional FDK methods because of the high degree of discomfort, and the thinness of the cranial structure makes it difficult to reconstruct a structurally intact skull without structural constraints in contrast methods. Compared with the baseline method, SPIDER has a more continuous bone structure in the reconstructed volume. The zoom-in figure of Figure~\ref{fig:recon} (a) shows the TMJ region. SPIDER demonstrates in addition to higher fidelity compared to other methods, especially in areas of clinical concern such as the TMJ, and is able to more realistically render a more accurate skeletal structure, makes SPIDER more valuable in clinical diagnosis of motor function and other tasks that require a more rigorous bone assemblage.\\


\begin{table*}[t]
\centering
\centering
\caption{Quantitative evaluation of the reconstructed CT volume from biplanar X-rays. The best performance is highlighted in bold.}
\label{tab:recon}
\renewcommand{\arraystretch}{1.2}
\small
\begin{tabular}{llccccc}
\toprule
\textbf{Method} & \textbf{Dataset} & \textbf{PSNR (dB) $\uparrow$} & \textbf{SSIM (\%)$\uparrow$} & \textbf{Dice (\%)$\uparrow$} & \textbf{HD95 (mm) $\downarrow$} & \textbf{CD (mm) $\downarrow$}  \\
\midrule
\multirow{2}{*}{FDK} 
    & Lucy              & $6.37 \pm 1.80$ & $0.05 \pm 0.148$ & - & - & - \\
    & TotalSegmentator  & $6.03 \pm 7.35$ & $0.06 \pm 0.265$ & - & -& - \\
\midrule
\multirow{2}{*}{X2CTGAN} 
    & Lucy              & $24.44 \pm 2.118$ & $0.78 \pm 0.110$ & $0.66 \pm 0.121$ & $\mathbf{2.43 \pm 3.321}$ & $\mathbf{1.75 \pm 3.145}$ \\
    & TotalSegmentator  & $20.77 \pm 2.474$ & $0.42 \pm 0.099$ & $0.41 \pm 0.174$ & $9.32 \pm 6.979$ & $8.49 \pm 6.507$ \\
\midrule
\multirow{2}{*}{PerX2CT} 
    & Lucy              & $23.83 \pm 1.056$ & $0.80 \pm 0.041$ & $0.60 \pm 0.134$ & $2.63 \pm 3.409$ & $2.02 \pm 3.904$ \\
    & TotalSegmentator  & $22.89 \pm 1.221$ & $0.63 \pm 0.062$ & $0.62 \pm 0.142$ & $4.44 \pm 2.040$ & $3.47 \pm 3.273$ \\
\midrule
\multirow{2}{*}{\textbf{SPIDER (Ours)}} 
    & Lucy              & $\mathbf{25.40 \pm 0.870}$ & $\mathbf{0.83 \pm 0.030}$ & $\mathbf{0.71 \pm 0.123}$ & $2.60 \pm 3.420$ & $1.88 \pm 4.133$ \\
    & TotalSegmentator  & $\mathbf{25.04 \pm 1.251}$ & $\mathbf{0.74 \pm 0.061}$ & $\mathbf{0.74 \pm 0.136}$ & $\mathbf{3.40 \pm 2.377}$ & $\mathbf{2.63 \pm 2.717}$ \\
\bottomrule
\end{tabular}

\end{table*}

Compared to the head CT in the LUCY dataset, the data in the totalSegmentator dataset has a more complex structure, containing complex bones (vertabrae, hip, sacrum etc.), muscle tissues (autochthon, gluteus, etc.), and organs (kidney etc.). Reconstructing a volume based on a biplanar with reasonable positional relationships and complete structure is more difficult in this dataset. As in the LUCY dataset, we joined all known segmentations for joint learning separately (actually containing 8 segmentations of the tissue structure).
The results are shown in Fig.~\ref{fig:recon}~(b). In this more difficult task, SPIDER gradually pulls away from the other three methods in terms of the angle of the reconstructed image. Because the reconstructed volume contains a large amount of structural information, but the intensity values of them are not very contrasting, this makes it more difficult for the comparison methods to reconstruct a volume that contains clear structural boundaries. However, when structural information is introduced, SIPIDER shows excellent reconstruction ability in which bones, organs, and muscles are clearly visible with distinct boundaries.

The quantitative metrics of all methods evaluated in both datasets are shown in Table~\ref{tab:recon}. The quality of SPIDER reconstructed volume is optimal in both datasets in terms of SSIM and PSNR metrics.

\subsection{Comparision of quality of reconstructed structures}

\begin{figure*}[ht]
    \centering
    \includegraphics[width=1\linewidth]{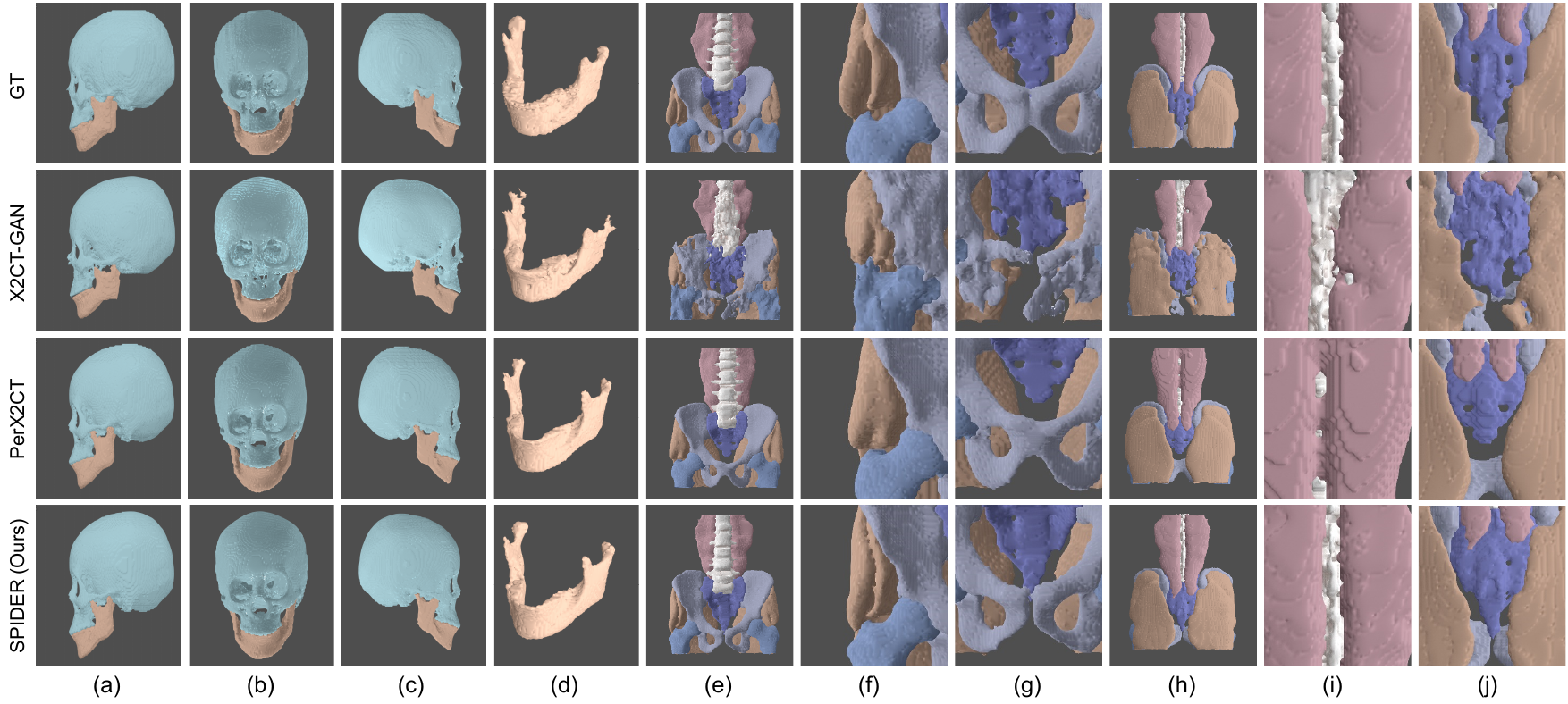}
    \caption{Comparison of the results of different methods for reconstructing the structural information in the volume, shows the mesh of different methods obtained using marchingcube with the same Laplace smoothing. (a)-(d) shows the result under LUCY Dataset, we additionally showed the results for the mandible alone to demonstrate the reconstruction effect on the joint. (e)-(j) shows the result and zoom-in figure of the meshes under Total Segmentator Dataset. }
    \label{fig:seg}
\end{figure*}
\begin{table*}[t]
\centering
\caption{Quantitative evaluation of pre-trained nnUNet segmentation downstream tasks in Total Segmentator Dataset with evaluation metrics demonstrating the quality of tissues and organs in the actual reconstructed volume, Optimal values in the table are marked in red.}
\label{tab:all_kinds_of_segmentations}
\renewcommand{\arraystretch}{1.2}
\footnotesize
\setlength{\tabcolsep}{3pt}
\begin{tabular}{l|ccc|ccc|ccc}
\toprule
\textbf{Structure} & \multicolumn{3}{c|}{\textbf{X2CT-GAN}} & \multicolumn{3}{c|}{\textbf{PerX2CT}} & \multicolumn{3}{c}{\textbf{SPIDER (Ours)}} \\
& DICE (\%) $\uparrow$ & HD95 (mm) $\downarrow$ & CD (mm) $\downarrow$ & DICE (\%) $\uparrow$ & HD95 (mm) $\downarrow$ & CD (mm) $\downarrow$ & DICE (\%) $\uparrow$ & HD95 (mm) $\downarrow$ & CD (mm) $\downarrow$ \\
\midrule
Maxilla      & $0.69 \pm 0.12$ & $\textcolor{blue}{2.31 \pm 3.05}$ & $\textcolor{blue}{1.65 \pm 2.47}$ & $0.65 \pm 0.12$ & $2.37 \pm 3.01$ & $1.73 \pm 2.30$ & $\textcolor{blue}{0.71 \pm 0.12}$ & $2.55 \pm 3.05$ & $1.80 \pm 2.53$ \\
Mandible     & $0.63 \pm 0.12$ & $\textcolor{blue}{2.54 \pm 3.61}$ & $\textcolor{blue}{1.85 \pm 3.74}$ & $0.54 \pm 0.12$ & $2.89 \pm 3.79$ & $2.31 \pm 5.05$ & $\textcolor{blue}{0.71 \pm 0.13}$ & $2.65 \pm 3.79$ & $1.97 \pm 5.31$ \\
Kidney       & $0.19 \pm 0.13$ & $18.2 \pm 12.2$ & $15.3 \pm 7.57$ & $0.44 \pm 0.13$ & $7.50 \pm 2.38$ & $7.41 \pm 4.33$ & $\textcolor{blue}{0.54 \pm 0.15}$ & $\textcolor{blue}{7.11 \pm 3.48}$ & $\textcolor{blue}{6.81 \pm 4.62}$ \\
Hip          & $0.46 \pm 0.11$ & $7.87 \pm 2.33$ & $5.64 \pm 3.94$ & $0.65 \pm 0.09$ & $3.76 \pm 1.44$ & $2.28 \pm 1.12$ & $\textcolor{blue}{0.77 \pm 0.07}$ & $\textcolor{blue}{2.73 \pm 1.50}$ & $\textcolor{blue}{1.42 \pm 0.61}$ \\
Sacrum       & $0.32 \pm 0.15$ & $6.94 \pm 4.78$ & $7.37 \pm 5.67$ & $0.58 \pm 0.12$ & $4.13 \pm 1.58$ & $2.36 \pm 1.27$ & $\textcolor{blue}{0.73 \pm 0.06}$ & $\textcolor{blue}{2.66 \pm 0.95}$ & $\textcolor{blue}{1.61 \pm 0.86}$ \\
Femur        & $0.42 \pm 0.17$ & $10.2 \pm 4.97$ & $10.5 \pm 7.22$ & $0.69 \pm 0.14$ & $4.16 \pm 1.74$ & $3.43 \pm 4.37$ & $\textcolor{blue}{0.81 \pm 0.17}$ & $\textcolor{blue}{3.01 \pm 2.42}$ & $\textcolor{blue}{1.76 \pm 1.65}$ \\
Vertebrae    & $0.46 \pm 0.09$ & $6.71 \pm 4.27$ & $6.78 \pm 5.20$ & $0.54 \pm 0.08$ & $4.15 \pm 1.41$ & $3.25 \pm 3.24$ & $\textcolor{blue}{0.75 \pm 0.07}$ & $\textcolor{blue}{2.56 \pm 1.01}$ & $\textcolor{blue}{2.13 \pm 1.48}$ \\
Autochthon   & $0.50 \pm 0.15$ & $8.02 \pm 3.95$ & $7.12 \pm 6.10$ & $0.74 \pm 0.09$ & $3.25 \pm 0.71$ & $\textcolor{blue}{2.16 \pm 1.03}$ & $\textcolor{blue}{0.81 \pm 0.09}$ & $\textcolor{blue}{2.68 \pm 1.01}$ & $2.32 \pm 1.57$ \\
Gluteus      & $0.56 \pm 0.10$ & $7.31 \pm 3.40$ & $6.68 \pm 3.54$ & $0.69 \pm 0.09$ & $4.11 \pm 1.47$ & $3.38 \pm 2.04$ & $\textcolor{blue}{0.79 \pm 0.06}$ & $\textcolor{blue}{3.07 \pm 0.84}$ & $\textcolor{blue}{2.35 \pm 1.18}$ \\
\midrule
\textbf{Avg} & $0.41 \pm 0.17$ & $9.32 \pm 6.98$ & $8.49 \pm 6.51$ & $0.62 \pm 0.14$ & $4.44 \pm 2.04$ & $3.47 \pm 3.27$ & $\textcolor{blue}{0.74 \pm 0.14}$ & $\textcolor{blue}{3.40 \pm 2.38}$ & $\textcolor{blue}{2.63 \pm 2.72}$ \\
\bottomrule
\end{tabular}
\end{table*}
To further evaluate the anatomical accuracy of the reconstructed CT volumes, we employ a pretrained segmentation network to extract organ, bone, and muscle structures from the reconstructed outputs of different models, then transformed them into meshes for display using the marchingcube algorithm with the same smoothing parameters. Fig.~\ref{fig:seg} presents the visual comparison among the ground truth CT, X2CT-GAN, PerX2CT, and our proposed method, SPIDER.

From the results, it is evident that SPIDER delivers the most faithful tissue reconstruction, producing segmentation results that closely resemble the ground truth in both shape and location. In contrast, the reconstructions from X2CT-GAN and PerX2CT often suffer from anatomical distortions, such as blurred boundaries or missing structures, particularly in regions involving soft tissues or complex interfaces between bones and muscles.\\
From the zoom-in plot Fig.~\ref{fig:seg}~(g), we can see that among the reconstructed volumes, only SPIDER reconstructs the hip bone boundary that can be easily recognized by the downstream segmentation network, whereas the pubic arch region cannot be reconstructed with correct separation. We can also see this phenomenon in some muscles, such as the autochthon shown in the zoom-in figure, where only our SPIDRER reconstructed volume can provide a clear volume for the downstream task to segment the two muscles accurately and completely. At the same time, we can also guarantee the correctness of the structure in the reconstructed volume. Taking the sacrum in the figure~\ref{fig:seg}~(j) as an example, the reconstructed volume from our SPIDRER network is much closer to the real reconstruction.\\
Table~\ref{tab:all_kinds_of_segmentations} calculates the metrics in the segmentation task targeting the skeletons of different organ tissues further confirming the phenomenon we observed. Combining the calculated results, we can understand that the SPIDER reconstructed volume contains more clinically valuable structural information, especially in the spine, which is a complex skeleton, and the kidney, which is a poorly contrasted in CT volumes.\\
The superior performance of SPIDER is attributed to its ability to better preserve spatial features and anatomical consistency during reconstruction. This ensures that when the same segmentation model is applied, the predicted body tissue masks from SPIDER-reconstructed CTs maintain clear tissue boundaries and structurally complete representations. These results demonstrate that SPIDER not only achieves visually realistic reconstruction but also retains critical clinical details necessary for downstream analysis, such as organ and tissue segmentation.

\subsection{Ablation Study}
\begin{figure*}[ht]
    \centering
    \includegraphics[width=1\linewidth]{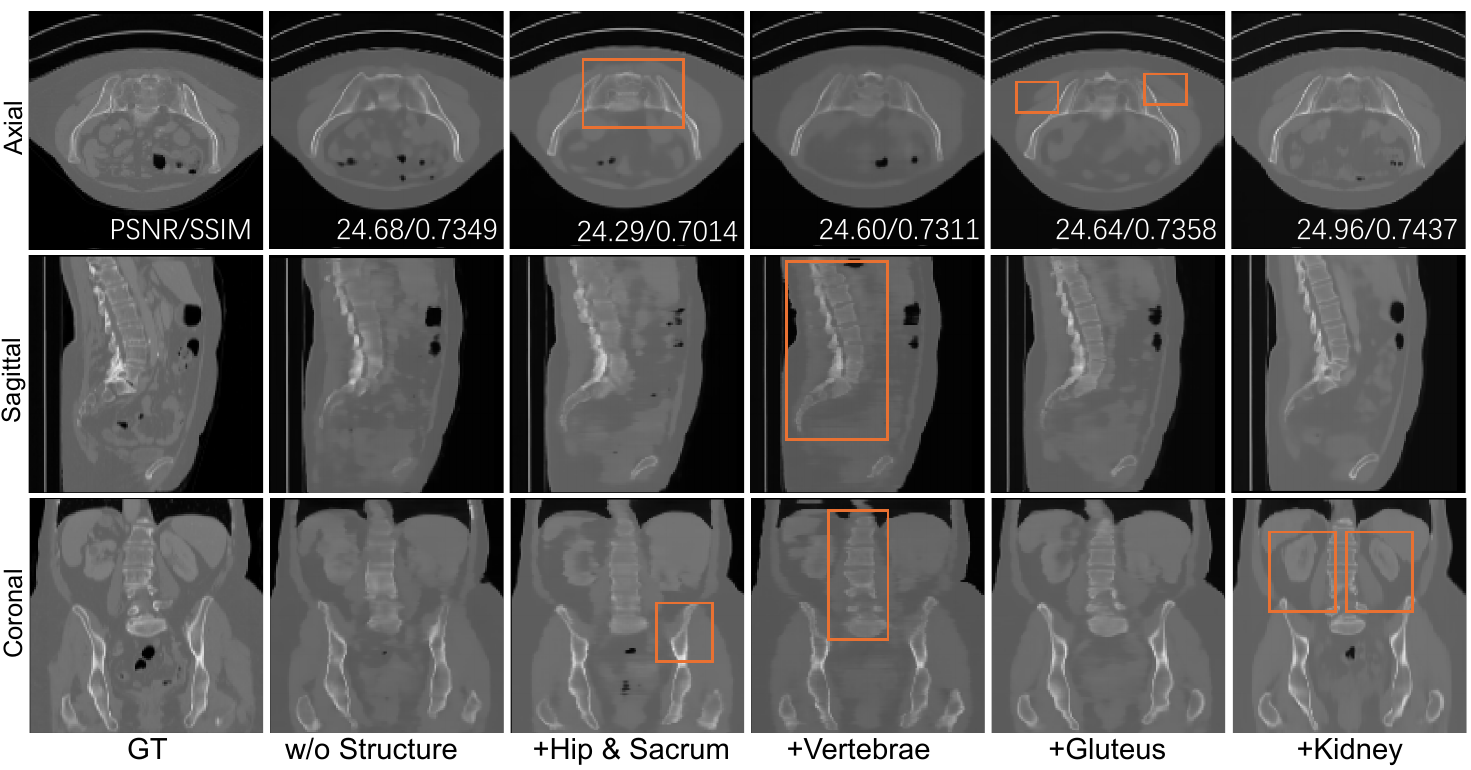}
    \caption{Reconstruction results of incrementally adding structural information are shown. The corresponding positions of the added structural information are boxed in the figure.}
    \label{fig:ablation_recon}
\end{figure*}
\begin{figure}[ht]
    \centering
    \includegraphics[width=1\linewidth]{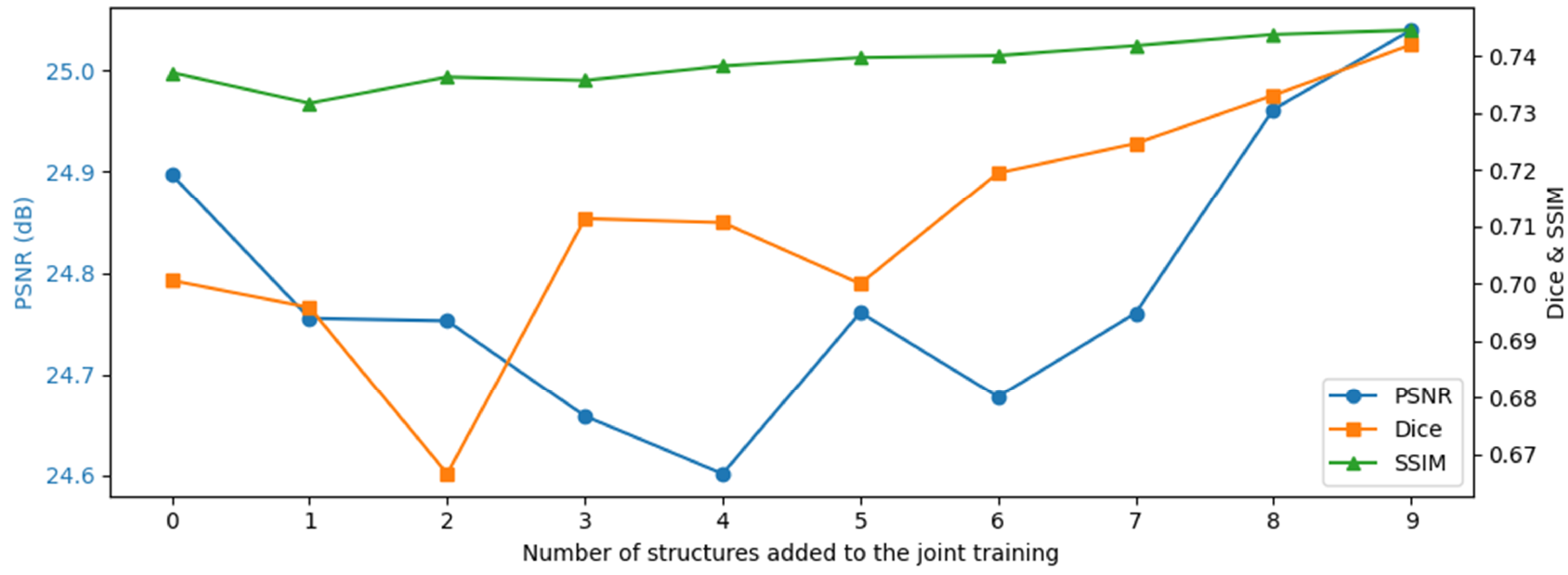}
    \caption{Reconstruction result evaluation metrics change with the asymptotic addition of structural information.}
    \label{fig:ablation_metrix}
\end{figure}
\subsubsection{Ablation experiments for joint training}
Here, we discuss the role that providing structural constraints in INR-based methods, i.e., the inclusion of structural information segmentation in SPIDER for joint training, has in reconstructing the complete structural recovery in an image. We demonstrate how this approach in SPIDER can enhance the reconstructed volume using a more complex Total Segmentator dataset. We start by training the volume alone, and use a gradual increase in the variety of segmentation until the segmentation containing structural information covers the most of the reconstructed volume was observed to observe the structural gain that adding a particular segmentation brought to the organs, bones, muscles, etc. represented by that segmentation in the volume.\\
Result is shown in Figure~\ref{fig:ablation_recon}, the boxed portion of the figure highlights the impact of the results resulting from the addition of segmentation joint training. When no structural information was added, the bones in the reconstructed images could be easily reconstructed but did not have distinct boundaries because of their high contrast, while the muscles with lower contrast were more difficult to distinguish from the organ tissues. As we increased the segmentation joint training for some of the bones (hip, sacrum), the reconstruction quality of these two parts of the bones was significantly enhanced, but the vertebrae remained blurred until their segmentation joint training was increased. Further, when we added gluteus, this muscle began to have a correct anatomical structure with a boundary between the gluteus medius and gluteus maximus. When we added the segmentation of the kidneys, the resulting reconstructed image also showed kidneys with distinct boundaries. As the proportion of volume accounted for by segmentation increased, the quantification of the reconstructed image gradually surpassed that of the joint training without the addition of segmentation. This demonstrates that the addition of segmented structural information to SPIDER has a significant effect on the structural integrity of the corresponding reconstructed image.\\

\begin{figure*}[ht]
    \centering
    \includegraphics[width=1\linewidth]{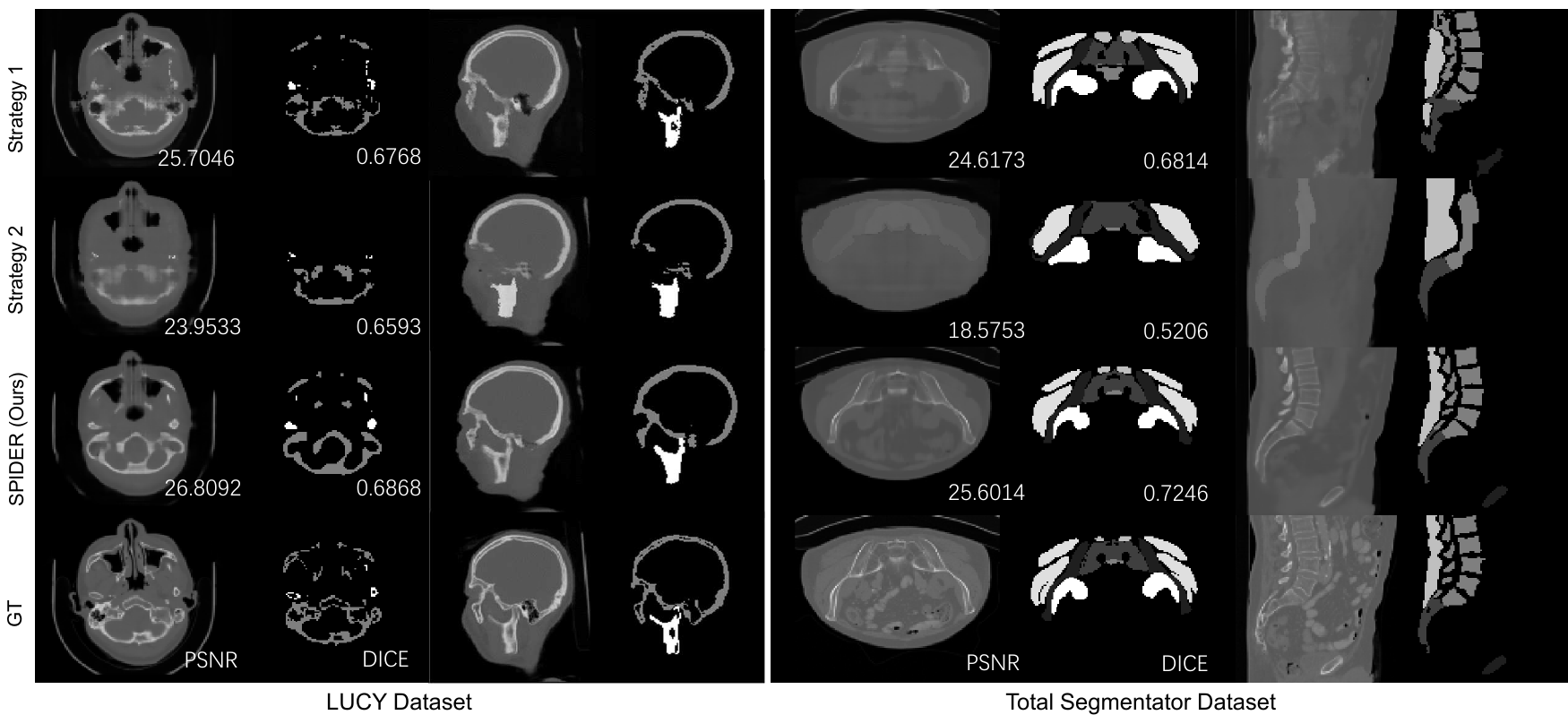}
    \caption{Comparison of the reconstructed volumes using SPIDER with different types of decoders with the results obtained using a pre-trained segmentation model as a downstream task.}
    \label{fig:stages_and_branches_result}
\end{figure*}
\begin{table*}[t]

\centering
\caption{Comparison of evaluation metrics of results reconstructed by SPIDER using different types of decoders.}
\label{tab:stages_and_branches_result}
\renewcommand{\arraystretch}{1.2}
\small
\begin{tabular}{llccccc}
\toprule
\textbf{Method} & \textbf{Dataset} & \textbf{PSNR (dB) $\uparrow$} & \textbf{SSIM (\%)$\uparrow$} & \textbf{Dice (\%)$\uparrow$} & \textbf{HD95 (mm) $\downarrow$} & \textbf{CD (mm) $\downarrow$} \\
\midrule
\multirow{2}{*}{SPIDER (a)} 
    & Lucy              & $25.31 \pm 0.989$ & $\mathbf{0.85 \pm 0.035}$ & $0.69 \pm 0.097$ & $2.62 \pm 3.505$ & $\mathbf{1.84 \pm 2.414}$ \\
    & TotalSegmentator  & $24.19 \pm 1.159$ & $0.66 \pm 0.084$ & $0.56 \pm 0.196$ & $6.84 \pm 6.897$ & $5.82 \pm 5.889$ \\

\midrule
\multirow{2}{*}{SPIDER (b)} 
    & Lucy              & $24.82 \pm 0.906$ & $0.82 \pm 0.034$ & $0.68 \pm 0.120$ & $5.03 \pm 4.023$ & $3.96 \pm 4.010$ \\
    & TotalSegmentator  & $23.19 \pm 1.029$ & $0.67 \pm 0.058$ & $0.34 \pm 0.252$ & $17.81\pm 12.488$ & $13.17 \pm 12.621$ \\
\midrule
\multirow{2}{*}{\textbf{SPIDER (Ours)}} 
    & Lucy              & $\mathbf{25.40 \pm 0.870}$ & $0.83 \pm 0.030$ & $\mathbf{0.71 \pm 0.123}$ & $\mathbf{2.60 \pm 3.420}$ & $1.88 \pm 4.133$ \\
    & TotalSegmentator  & $\mathbf{25.04 \pm 1.251}$ & $\mathbf{0.74 \pm 0.061}$ & $\mathbf{0.74 \pm 0.136}$ & $\mathbf{3.40 \pm 2.377}$ & $\mathbf{2.63 \pm 2.717}$ \\
\bottomrule
\end{tabular}

\end{table*}
 \subsubsection{Ablation experiments for shared decoders}

In SPIDER, we chose the form of shared decoder and joint training of volume and segmentation to implicitly incorporate structural constraints into peer-to-peer prediction learning. In this section, to demonstrate the effectiveness of the shared decoder in SPIDER:
\begin{itemize}
    \item Strategy 1: two-branches. We use two MLP decoders identical to those in SPIDER to decode volume and segmentation, respectively, shown in Fig.~\ref{fig:branch};
     \begin{figure}[h]
    \centering
    \includegraphics[width=0.8\linewidth]{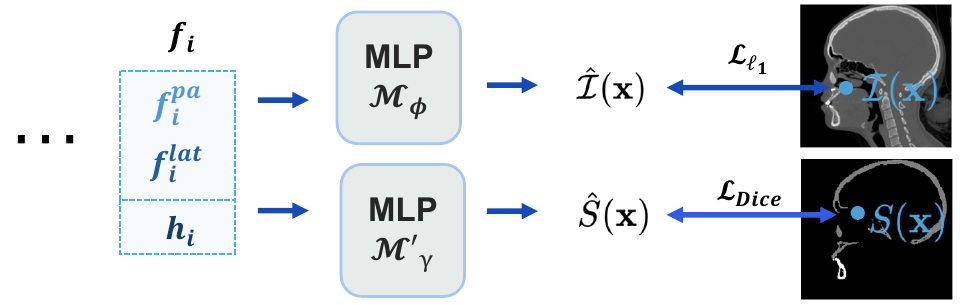}
    \caption{Network structure of a two-branches form decoder.}
    \label{fig:branch}
\end{figure}
    \item Strategy 2 two-stages : we first use one decoder to decode the volume, followed by one MLP to get the classification of the output points. Intuitively, structural decoding can be viewed as a reconstruction decoding with an adaptive threshold, implemented by MLP. This structure also does not share a decoder, but it establishes an explicit relationship between structural decoding and reconstruction decoding, shown in Fig.~\ref{fig:stage}
     \begin{figure}[h]
    \centering
    \includegraphics[width=0.8\linewidth]{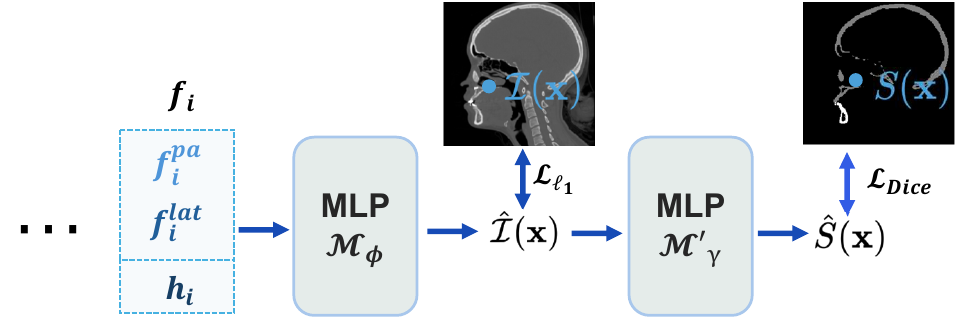}
    \caption{Network structure of a two-stages form decoder.}
    \label{fig:stage}
\end{figure}
\end{itemize}
A comparison of the results obtained using these two jointly trained decoder approaches with the approach in SPIDER is shown in Fig.~\ref{fig:stages_and_branches_result} Only the use of the shared decoder performs the best, replacing other forms of decoder will not give results with a clear structure. Table.~\ref{tab:stages_and_branches_result} also corroborates this on the evaluation metrics for the image and downstream segmentation tasks. This proves that joint training using a shared decoder is necessary in SPIDER.

\section{Discussion}
\subsection{Potential in Segmentation Tasks}
\begin{figure}[ht]
    \centering
    \includegraphics[width=1\linewidth]{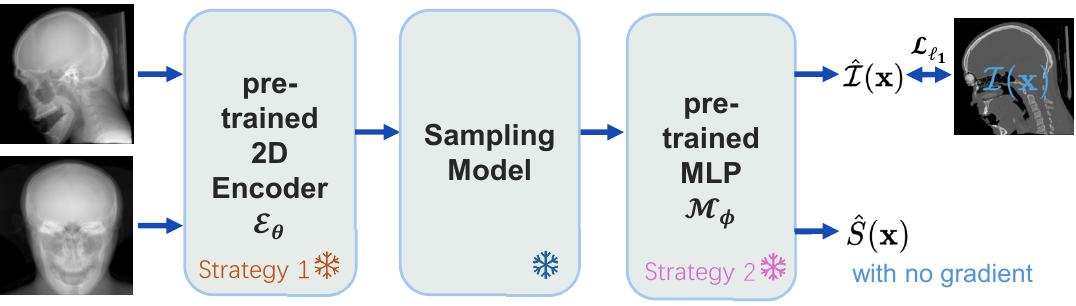}
    \caption{Pipeline of the pre-trained SPIDER model is used as an implicit segmentation model.}
    \label{fig:frozing}
\end{figure}

\begin{figure}[ht]
    \centering
    \includegraphics[width=1\linewidth]{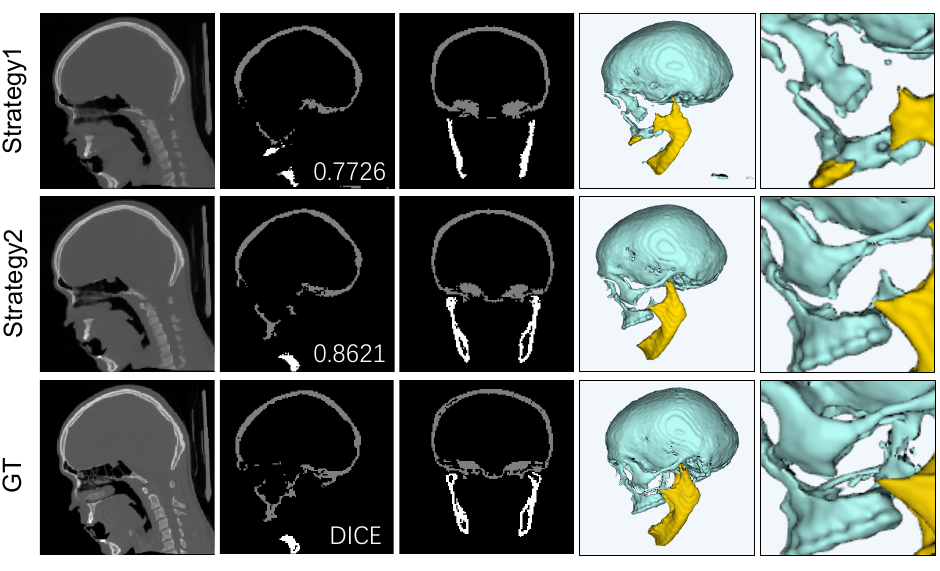}
    \caption{The result of two frozing strategy. The overfitted volume, predicted segmentation, and mesh obtained using marching cubes are all shown in the figure.}
    \label{fig:frozing_result}
\end{figure}

In previous experiments, it was found that SPIDER can be enhanced to reconstruct structural information in the volume by implicitly learning the relationship between point-to-point intensity and segmentation. As can be seen from the experiments in our ablation study, the implementation of this functionality relies heavily on shared decoder and jointly train. Based on this, we cleverly apply the pre-trained SPIDER to the segmentation task. Fig.~\ref{fig:frozing} illustrates our model design. Firstly, we used the data in the LUCY dataset (jaw-close data) to complete pre-training on the SPIDER network. Then, we use an out-of-domain data (jaw-open data) biplanar Xray of the same body part as input of the network, while freezing the decoder parameters, the gradient back propagation is done only at the loss of the intensity prediction, self-supervising the fitting of the intensity prediction to the real CT volume. Immediately after, we observe the segmentation of the MLP joint output as a segmentation of this extra-domain CT volume.
In practice, we used a simulated jaw-open head CT volume with the placement as known information to simulate its corresponding biplane Xray. To further verify that it is the shared decoder that learns the pixel-level intensity-structure mapping, we froze the pre-trained decoder's parameters as in case2 in Fig.~\ref{fig:frozing}. As a comparison, we also froze the pre-trained encoder to further validate the above conclusion.\\
The experimental results we obtained are shown in Fig.~\ref{fig:frozing_result}. Compared with the freezing encoder case, the result of freezing decoder case shows a better segmentation effect after overfitting the network to a simulated CT volume. Our pre-trained network on jaw-close data is able to learn segmentation indirectly by learning the real CT volume while shows that the SPIDER model learns the point-to-point classification relationship by the MLP decoder, which further verifies the conclusions proposed in the ablation experiments and further enhances the interpretability of the SPIDER model.

\subsection{Limitation}
SPIDER may cause some parts of the reconstruction outside the region of joining structural information to become worse and lack some high-frequency information. As shown in Fig.~\ref{fig:ablation_recon} and Fig.~\ref{fig:ablation_metrix}, this can lead to an overall degradation in the quality of the image if only a part of the structural information that accounts for a small portion of the entire VOLUME is added for joint training. However, as more structural information regions are added until the whole volume is covered, the quality of the image gradually increases. Also, SPIDER is sensitive to the weights of the joint training loss function, especially for datasets with a large variety of structural information, a lot of attempts are needed to get more suitable weights to train the network more smoothly and get better reconstruction results.

\section{Conclusion}
In this work, we proposed SPIDER, a Structure-Prior Implicit Deep Network, for the challenging task of reconstructing volumetric CT images from only two biplanar X-ray projections. Unlike traditional CNN- or GAN-based models, SPIDER embeds anatomical structural priors—such as tissue segmentation—directly into an implicit neural representation decoder by jointly training. This training approach enables the decoder to learn the implicit pixel to pixel mapping of intensity and structure information.

Through extensive experiments on two datasets with varying anatomical complexity, we demonstrated that SPIDER outperforms state-of-the-art methods like X2CT-GAN and PerX2CT in both image quality and structural fidelity. Notably, the reconstructed volumes from SPIDER show enhanced anatomical continuity and clearer tissue boundaries, especially in clinically important regions such as the temporomandibular joint and more difficult challenges ,such as reconstructing the tissue of Total Segmentator Dataset. Downstream segmentation tasks further verify the superior visibility and positional accuracy of bones, muscles, and organs in SPIDER's outputs.

Moreover, SPIDER exhibits strong generalization to complex anatomical patterns and low-contrast regions, making it a promising tool for real-world applications like surgical planning and diagnostic support. In summary, our structure-prior implicit representation framework offers a practical and effective solution for CT reconstruction from biplanar X-rays, bridging the gap between data efficiency and clinical fidelity in sparse-view medical imaging.

In summary, SPIDER provides a new way of thinking to ensure the structural stability of very underdetermined CT reconstruction methods based on implicit representations.

\bibliographystyle{IEEEtran}  
\bibliography{ref}

\begin{thebibliography}{10}
\providecommand{\url}[1]{#1}
\csname url@samestyle\endcsname
\providecommand{\newblock}{\relax}
\providecommand{\bibinfo}[2]{#2}
\providecommand{\BIBentrySTDinterwordspacing}{\spaceskip=0pt\relax}
\providecommand{\BIBentryALTinterwordstretchfactor}{4}
\providecommand{\BIBentryALTinterwordspacing}{\spaceskip=\fontdimen2\font plus
\BIBentryALTinterwordstretchfactor\fontdimen3\font minus \fontdimen4\font\relax}
\providecommand{\BIBforeignlanguage}[2]{{%
\expandafter\ifx\csname l@#1\endcsname\relax
\typeout{** WARNING: IEEEtran.bst: No hyphenation pattern has been}%
\typeout{** loaded for the language `#1'. Using the pattern for}%
\typeout{** the default language instead.}%
\else
\language=\csname l@#1\endcsname
\fi
#2}}
\providecommand{\BIBdecl}{\relax}
\BIBdecl

\bibitem{kalender2011computed}
W.~A. Kalender, \emph{Computed tomography: fundamentals, system technology, image quality, applications}.\hskip 1em plus 0.5em minus 0.4em\relax John Wiley \& Sons, 2011.

\bibitem{hsieh2003computed}
J.~Hsieh, ``Computed tomography: principles, design, artifacts, and recent advances,'' 2003.

\bibitem{brenner2007computed}
D.~J. Brenner and E.~J. Hall, ``Computed tomography—an increasing source of radiation exposure,'' \emph{New England journal of medicine}, vol. 357, no.~22, pp. 2277--2284, 2007.

\bibitem{loisel2023three}
F.~Loisel, S.~Durand, J.-N. Goubier, X.~Bonnet, P.~Rouch, and W.~Skalli, ``Three-dimensional reconstruction of the hand from biplanar x-rays: Assessment of accuracy and reliability,'' \emph{Orthopaedics \& Traumatology: Surgery \& Research}, vol. 109, no.~6, p. 103403, 2023.

\bibitem{yang20242d}
D.~Yang, H.~Shi, B.~Zeng, and X.~Chen, ``2d/3d registration based on biplanar x-ray and ct images for surgical navigation,'' \emph{Computer Methods and Programs in Biomedicine}, vol. 257, p. 108444, 2024.

\bibitem{kadoury2008statistical}
S.~Kadoury, F.~Cheriet, and H.~Labelle, ``A statistical image-based approach for the 3d reconstruction of the scoliotic spine from biplanar radiographs,'' in \emph{2008 5th IEEE International Symposium on Biomedical Imaging: From Nano to Macro}.\hskip 1em plus 0.5em minus 0.4em\relax IEEE, 2008, pp. 660--663.

\bibitem{humbert20093d}
L.~Humbert, J.~A. De~Guise, B.~Aubert, B.~Godbout, and W.~Skalli, ``3d reconstruction of the spine from biplanar x-rays using parametric models based on transversal and longitudinal inferences,'' \emph{Medical engineering \& physics}, vol.~31, no.~6, pp. 681--687, 2009.

\bibitem{wasserthal2023totalsegmentator}
J.~Wasserthal, H.-C. Breit, M.~T. Meyer, M.~Pradella, D.~Hinck, A.~W. Sauter, T.~Heye, D.~T. Boll, J.~Cyriac, S.~Yang \emph{et~al.}, ``Totalsegmentator: robust segmentation of 104 anatomic structures in ct images,'' \emph{Radiology: Artificial Intelligence}, vol.~5, no.~5, p. e230024, 2023.

\bibitem{liu2021deep}
P.~Liu, H.~Han, Y.~Du, H.~Zhu, Y.~Li, F.~Gu, H.~Xiao, J.~Li, C.~Zhao, L.~Xiao \emph{et~al.}, ``Deep learning to segment pelvic bones: large-scale ct datasets and baseline models,'' \emph{International Journal of Computer Assisted Radiology and Surgery}, vol.~16, pp. 749--756, 2021.

\bibitem{sekuboyina2021verse}
A.~Sekuboyina, M.~E. Husseini, A.~Bayat, M.~L{\"o}ffler, H.~Liebl, H.~Li, G.~Tetteh, J.~Kuka{\v{c}}ka, C.~Payer, D.~{\v{S}}tern \emph{et~al.}, ``Verse: a vertebrae labelling and segmentation benchmark for multi-detector ct images,'' \emph{Medical image analysis}, vol.~73, p. 102166, 2021.

\bibitem{lin2023rsna}
H.~M. Lin, E.~Colak, T.~Richards, F.~C. Kitamura, L.~M. Prevedello, J.~Talbott, R.~L. Ball, E.~Gumeler, K.~W. Yeom, M.~Hamghalam \emph{et~al.}, ``The rsna cervical spine fracture ct dataset,'' \emph{Radiology: Artificial Intelligence}, vol.~5, no.~5, p. e230034, 2023.

\bibitem{armato2011lung}
S.~G. Armato~III, G.~McLennan, L.~Bidaut, M.~F. McNitt-Gray, C.~R. Meyer, A.~P. Reeves, B.~Zhao, D.~R. Aberle, C.~I. Henschke, E.~A. Hoffman \emph{et~al.}, ``The lung image database consortium (lidc) and image database resource initiative (idri): a completed reference database of lung nodules on ct scans,'' \emph{Medical physics}, vol.~38, no.~2, pp. 915--931, 2011.

\bibitem{X-CTRSNet}
R.~Ge, Y.~He, C.~Xia, C.~Xu, W.~Sun, G.~Yang, J.~Li, Z.~Wang, H.~Yu, D.~Zhang \emph{et~al.}, ``X-ctrsnet: 3d cervical vertebra ct reconstruction and segmentation directly from 2d x-ray images,'' \emph{Knowledge-Based Systems}, vol. 236, p. 107680, 2022.

\bibitem{X-ray2CTNet}
X.~Sun, X.~Li, and P.~Chen, ``An ultra-sparse view ct imaging method based on x-ray2ctnet,'' \emph{IEEE Transactions on Computational Imaging}, vol.~8, pp. 733--742, 2022.

\bibitem{X2ctgan}
X.~Ying, H.~Guo, K.~Ma, J.~Wu, Z.~Weng, and Y.~Zheng, ``X2ct-gan: reconstructing ct from biplanar x-rays with generative adversarial networks,'' in \emph{Proceedings of the IEEE/CVF conference on computer vision and pattern recognition}, 2019, pp. 10\,619--10\,628.

\bibitem{Xtransct}
C.~Zhang, L.~Liu, J.~Dai, X.~Liu, W.~He, Y.~Chan, Y.~Xie, F.~Chi, and X.~Liang, ``Xtransct: ultra-fast volumetric ct reconstruction using two orthogonal x-ray projections for image-guided radiation therapy via a transformer network,'' \emph{Physics in Medicine and Biology}, vol.~69, no.~8, p. 085010, 2024.

\bibitem{CCX}
M.~A.~R. Ratul, K.~Yuan, and W.~Lee, ``Ccx-raynet: a class conditioned convolutional neural network for biplanar x-rays to ct volume,'' in \emph{2021 IEEE 18th International Symposium on Biomedical Imaging (ISBI)}.\hskip 1em plus 0.5em minus 0.4em\relax IEEE, 2021, pp. 1655--1659.

\bibitem{wang2018esrgan}
X.~Wang, K.~Yu, S.~Wu, J.~Gu, Y.~Liu, C.~Dong, Y.~Qiao, and C.~Change~Loy, ``Esrgan: Enhanced super-resolution generative adversarial networks,'' in \emph{Proceedings of the European conference on computer vision (ECCV) workshops}, 2018, pp. 0--0.

\bibitem{DIFNet}
Y.~Lin, Z.~Luo, W.~Zhao, and X.~Li, ``Learning deep intensity field for extremely sparse-view cbct reconstruction,'' in \emph{International Conference on Medical Image Computing and Computer-Assisted Intervention}.\hskip 1em plus 0.5em minus 0.4em\relax Springer, 2023, pp. 13--23.

\bibitem{kyung2023perspective}
D.~Kyung, K.~Jo, J.~Choo, J.~Lee, and E.~Choi, ``Perspective projection-based 3d ct reconstruction from biplanar x-rays,'' in \emph{ICASSP 2023-2023 IEEE International Conference on Acoustics, Speech and Signal Processing (ICASSP)}.\hskip 1em plus 0.5em minus 0.4em\relax IEEE, 2023, pp. 1--5.

\bibitem{X2ctCNN}
Y.~Sun, T.~Netherton, L.~Court, A.~Veeraraghavan, and G.~Balakrishnan, ``Ct reconstruction from few planar x-rays with application towards low-resource radiotherapy,'' in \emph{International Conference on Medical Image Computing and Computer-Assisted Intervention}.\hskip 1em plus 0.5em minus 0.4em\relax Springer, 2023, pp. 225--234.

\bibitem{borzabadi2012orthodontic}
A.~Borzabadi-Farahani, ``Orthodontic considerations in restorative management of hypodontia patients with endosseous implants,'' \emph{Journal of Oral Implantology}, vol.~38, no.~6, pp. 779--791, 2012.

\bibitem{schmalbach2010anterior}
C.~E. Schmalbach, D.~E. Webb, and E.~K. Weitzel, ``Anterior skull base reconstruction: a review of current techniques,'' \emph{Current opinion in otolaryngology \& head and neck surgery}, vol.~18, no.~4, pp. 238--243, 2010.

\bibitem{zuniga2016updates}
M.~G. Zuniga, J.~H. Turner, and R.~K. Chandra, ``Updates in anterior skull base reconstruction,'' \emph{Current Opinion in Otolaryngology \& Head and Neck Surgery}, vol.~24, no.~1, pp. 75--82, 2016.

\bibitem{lee2023locality}
D.~Lee, C.~Kim, M.~Cho, and W.~S. HAN, ``Locality-aware generalizable implicit neural representation,'' \emph{Advances in Neural Information Processing Systems}, vol.~36, pp. 48\,363--48\,381, 2023.

\bibitem{ekanayake2025seco}
M.~Ekanayake, Z.~Chen, G.~Egan, M.~Harandi, and Z.~Chen, ``Seco-inr: Semantically conditioned implicit neural representations for improved medical image super-resolution,'' in \emph{2025 IEEE/CVF Winter Conference on Applications of Computer Vision (WACV)}.\hskip 1em plus 0.5em minus 0.4em\relax IEEE, 2025, pp. 117--126.

\bibitem{zhang2024attention}
S.~Zhang, K.~Liu, J.~Gu, X.~Cai, Z.~Wang, J.~Bu, and H.~Wang, ``Attention beats linear for fast implicit neural representation generation,'' in \emph{European Conference on Computer Vision}.\hskip 1em plus 0.5em minus 0.4em\relax Springer, 2024, pp. 1--18.

\bibitem{ronneberger2015u}
O.~Ronneberger, P.~Fischer, and T.~Brox, ``U-net: Convolutional networks for biomedical image segmentation,'' in \emph{Medical image computing and computer-assisted intervention--MICCAI 2015: 18th international conference, Munich, Germany, October 5-9, 2015, proceedings, part III 18}.\hskip 1em plus 0.5em minus 0.4em\relax Springer, 2015, pp. 234--241.

\bibitem{muller2022instant}
T.~M{\"u}ller, A.~Evans, C.~Schied, and A.~Keller, ``Instant neural graphics primitives with a multiresolution hash encoding,'' \emph{ACM transactions on graphics (TOG)}, vol.~41, no.~4, pp. 1--15, 2022.

\bibitem{milletari2016v}
F.~Milletari, N.~Navab, and S.-A. Ahmadi, ``V-net: Fully convolutional neural networks for volumetric medical image segmentation,'' in \emph{2016 fourth international conference on 3D vision (3DV)}.\hskip 1em plus 0.5em minus 0.4em\relax Ieee, 2016, pp. 565--571.

\bibitem{Lucy}
Z.~Qiu, Y.~Li, D.~He, Q.~Zhang, L.~Zhang, Y.~Zhang, J.~Wang, L.~Xu, X.~Wang, Y.~Zhang \emph{et~al.}, ``Sculptor: Skeleton-consistent face creation using a learned parametric generator,'' \emph{ACM Transactions on Graphics (TOG)}, vol.~41, no.~6, pp. 1--17, 2022.

\bibitem{totalsegmentator}
J.~Wasserthal, H.-C. Breit, M.~T. Meyer, M.~Pradella, D.~Hinck, A.~W. Sauter, T.~Heye, D.~T. Boll, J.~Cyriac, S.~Yang, M.~Bach, and M.~Segeroth, ``Totalsegmentator: Robust segmentation of 104 anatomic structures in ct images,'' \emph{Radiology: Artificial Intelligence}, vol.~5, no.~5, p. e230024, 2023.

\bibitem{CILp1}
J.~S. J{\o}rgensen, E.~Ametova, G.~Burca, G.~Fardell, E.~Papoutsellis, E.~Pasca, K.~Thielemans, M.~Turner, R.~Warr, W.~R. Lionheart \emph{et~al.}, ``Core imaging library-part i: a versatile python framework for tomographic imaging,'' \emph{Philosophical Transactions of the Royal Society A}, vol. 379, no. 2204, p. 20200192, 2021.

\bibitem{CILp2}
E.~Papoutsellis, E.~Ametova, C.~Delplancke, G.~Fardell, J.~S. J{\o}rgensen, E.~Pasca, M.~Turner, R.~Warr, W.~R. Lionheart, and P.~J. Withers, ``Core imaging library-part ii: multichannel reconstruction for dynamic and spectral tomography,'' \emph{Philosophical Transactions of the Royal Society A}, vol. 379, no. 2204, p. 20200193, 2021.

\bibitem{FDK}
L.~A. Feldkamp, L.~C. Davis, and J.~W. Kress, ``Practical cone-beam algorithm,'' \emph{Josa a}, vol.~1, no.~6, pp. 612--619, 1984.

\bibitem{SSIM}
Z.~Wang, A.~C. Bovik, H.~R. Sheikh, and E.~P. Simoncelli, ``Image quality assessment: from error visibility to structural similarity,'' \emph{IEEE transactions on image processing}, vol.~13, no.~4, pp. 600--612, 2004.

\bibitem{isensee2021nnu}
F.~Isensee, P.~F. Jaeger, S.~A. Kohl, J.~Petersen, and K.~H. Maier-Hein, ``nnu-net: a self-configuring method for deep learning-based biomedical image segmentation,'' \emph{Nature methods}, vol.~18, no.~2, pp. 203--211, 2021.

\bibitem{Dice}
A.~Abdollahi, B.~Pradhan, and A.~Alamri, ``Vnet: An end-to-end fully convolutional neural network for road extraction from high-resolution remote sensing data,'' \emph{Ieee Access}, vol.~8, pp. 179\,424--179\,436, 2020.

\end{thebibliography}
\end{document}